\documentclass{article}
\usepackage{enumitem, amsmath,graphicx,url,lipsum}
\usepackage{tabularx,booktabs}
\usepackage{pifont}
\usepackage{amssymb}
\usepackage[toc,page]{appendix}
\usepackage[margin=1.0in]{geometry} 
\usepackage{listings}

\newcommand{\cmark}{\ding{51}}%
\newcommand{\xmark}{\ding{55}}%
\newcommand{\code}[1]{\texttt{#1}}

\title{The ICME 2025 Audio Encoder Capability Challenge}
\author{
Junbo Zhang$^{1}$\thanks{zhangjunbo1@xiaomi.com},
Heinrich Dinkel$^{1}$,
Qiong Song$^{2}$,
Helen Wang$^{2}$,
Yadong Niu$^{1}$,
Si Cheng$^{1}$, \\
Xiaofeng Xin$^{2}$,
Ke Li$^{2}$,
Wenwu Wang$^{3}$\thanks{w.wang@surrey.ac.uk},
Yujun Wang$^{1}$,
Jian Luan$^{1}$ \\
\small $^{1}$Xiaomi Corporation, China \quad 
\small $^{2}$Dataocean AI Inc., USA \quad  
\small $^{3}$University of Surrey, UK
}

\date{}

\begin{document}

\maketitle

\begin{abstract}
This challenge aims to evaluate the capabilities of audio encoders, especially in the context of multi-task learning and real-world applications.
Participants are invited to submit pre-trained audio encoders that map raw waveforms to continuous embeddings.
These encoders will be tested across diverse tasks including speech, environmental sounds, and music, with a focus on real-world usability. 
The challenge features two tracks: Track A for parameterized evaluation, and Track B for parameter-free evaluation.
This challenge provides a platform for evaluating and advancing the state-of-the-art in audio encoder design.
\end{abstract}

\section{Challenge Description}

The field of audio representation learning has advanced significantly in recent years, enabling models to extract meaningful features from audio data effectively. While much of the focus has been on discrete representations and tokenization \cite{van2017neural,mentzer2023finite,kumar2024high,siuzdak2024snac,kyutai2024moshi} recently, continuous representations remain crucial for many tasks. Unlike discrete tokens, continuous embeddings retain the nuanced information within audio signals, leading to better performance \cite{wang2024comparative} in downstream tasks like fine-grained classification, regression, and time-series analysis. Moreover, continuous audio encoders play a key role in multimodal large language models, facilitating the integration of audio with other modalities \cite{chu2023qwen,xie2024mini,yu2024salmonn}. Models such as wav2vec2 \cite{baevski2020wav2vec}, Data2vec2 \cite{baevski2022data2vec}, and Dasheng \cite{dinkel2024dasheng} have demonstrated strong performance across various audio tasks. While there are some existing benchmarks \cite{turian2022hear,yang2021superb,mousavi2024dasb} for evaluating these models, they leave room for further refinement, and a comparison with similar benchmarks can be found in Appendix A.

This challenge provides a platform for participants to showcase innovative approaches in model design and data utilization, pushing the boundaries of audio representation learning. Participants are required to submit a single pre-trained encoder that processes audio waveforms and generates two outputs: a sequence of continuous embedding vectors for frame-level tasks and a fixed-dimension embedding vector for utterance-level tasks. The model should comply with an API specified by the organizers, with examples provided\footnote{\url{https://github.com/jimbozhang/xares-template/blob/main/examples/dasheng/dasheng_encoder.py}}\textsuperscript{,}\footnote{\url{https://github.com/jimbozhang/xares-template/blob/main/examples/wav2vec2/wav2vec2_encoder.py}}.

The submitted models will be evaluated on diverse audio tasks, including human voice, environmental sounds, and music, using an open-source evaluation system\footnote{\url{https://github.com/jimbozhang/xares}}. Participants may test and optimize models independently, but the final rankings will be based on the evaluations by the organizers.

\begin{table*}[htb]
\centering
\caption{Datasets for Fine-tuning and evaluation. The hidden datasets marked by $^{\dagger}$. All shown tasks are evaluated in Track A. For Track B (unparameterized evaluation, see Section \ref{sec:tasks}), a subset of Track A's utterance-level classification tasks are selected. The preprocessed version of the datasets will be provided on Zenodo.}
\vspace{0.5em}
\label{tab:testset}
\begin{tabularx}{\textwidth}{Xlllrc}
\toprule
\textbf{Domain}      & \textbf{Dataset}                                             & \textbf{Task Type}                & \textbf{Metric}   & \textbf{\#} & \textbf{Track B} \\
\midrule
\textbf{Speech}      & Speech Commands    \cite{warden2018speech}                   & Keyword spotting                  & Acc               & 30                 & \cmark \\
                     & LibriCount \cite{stoter2018libricount}                       & Speaker counting                  & Acc               & 11                 & \cmark \\
                     & VoxLingua107 \cite{valk2021voxlingua107}                     & Language identification           & Acc               & 33                 & \cmark \\
                     & VoxCeleb1 \cite{nagrani2020voxceleb}                         & Speaker identification            & Acc               & 1251               & \cmark \\
                     & LibriSpeech \cite{panayotov2015librispeech}                  & Gender classification             & Acc               & 2                  & \cmark \\
                     & Fluent Speech Commands \cite{lugosch2019speech}              & Intent classification             & Acc               & 248                & \cmark \\
                     & VocalSound \cite{gong_vocalsound}                            & Non-speech sounds                 & Acc               & 6                  & \cmark \\
                     & CREMA-D \cite{cao2014crema}                                  & Emotion recognition               & Acc               & 5                  & \cmark \\
                     & speechocean762 \cite{zhang2021speechocean762}                & Phoneme pronunciation             & MSE               & 3                  & \xmark \\
                     & ASV2015 \cite{kinnunen2018automatic}                         & Spoofing detection                & EER               & 2                  & \cmark \\
\midrule                
\textbf{Sound}       & ESC-50 \cite{piczak2015esc}                                  & Environment classification        & Acc               & 50                 & \cmark \\
                     & FSD50k \cite{fonseca2021fsd50k}                              & Sound event detection             & mAP               & 200                & \xmark \\
                     & UrbanSound 8k \cite{salamon2014dataset}                      & Urban sound classification        & Acc               & 10                 & \cmark \\
                     & DESED \cite{turpault2019sound}                               & Sound event detection             & Segment-F1        & 10                 & \cmark \\
                     & FSD18-Kaggle \cite{fonseca2018general}                       & Sound event detection             & mAP               & 41                 & \xmark \\
                     & Clotho \cite{drossos2020clotho}                              & Sound retrieval                   & Recall@1          & -                  & \xmark \\
                     & Inside/outside car$^{\dagger}$                               & Sound event detection             & Acc               & 2                  & \cmark \\
                     & Finger snap sound$^{\dagger}$                                & Sound event detection             & Acc               & 2                  & \cmark \\
                     & Key scratching car$^{\dagger}$                               & Sound event detection             & Acc               & 2                  & \cmark \\
                     & Subway broadcast$^{\dagger}$                                 & Sound event detection             & Acc               & 2                  & \cmark \\
                     & LiveEnv sounds$^{\dagger}$                                   & Sound event detection             & mAP               & 18                 & \xmark \\
\midrule                
\textbf{Music}       & MAESTRO \cite{hawthorne2018enabling}                         & Note classification               & Acc               & 88                 & \cmark \\
                     & GTZAN Genre \cite{sturm2013gtzan}                            & Genre classification              & Acc               & 10                 & \cmark \\
                     & NSynth-Instruments \cite{nsynth2017}                         & Instruments Classification        & Acc               & 11                 & \cmark \\
                     & NSynth-Pitch \cite{nsynth2017}                               & Pitches Classification            & Acc               & 128                & \cmark \\
                     & Free Music Archive Small \cite{defferrard2016fma}            & Music genre classification        & Acc               & 8                  & \cmark \\
\bottomrule
\end{tabularx}
\end{table*}

\begin{table}[tb]
\centering
\caption{The hidden datasets provided by the Challenge organizers. These datasets are concealed from participants.}
\vspace{0.5em}
\label{tab:organizer_datasets}
\begin{tabular}{lllc}
\toprule
\textbf{Dataset}           & \textbf{Size}  & \textbf{Description}\\
\midrule
Inside/outside car         & 15k samples    & Inside car or outside car, for security threat prevention \\
Finger snap sound          & 15k samples    & Wake-up word alternative for smart speakers               \\
Key scratching car         & 5k samples     & Car vandalism through key scratching                      \\
Subway broadcast           & 125 hours      & Subway announcements broadcasting                         \\
LiveEnv sound              & 25k samples    & Environmental sounds recorded from real scenarios         \\
\bottomrule    
\end{tabular}
\end{table}

\subsection{Tracks}
\label{sec:tasks}

The challenge consists of two tracks, each evaluating the pre-trained models in different ways.

\textbf{Track A: Linear Fine-Tuning on Task-Specific Data.} A linear layer will be trained using the provided user embeddings, optimized with predefined hyperparameters for each task.
This approach assesses how effectively the fixed representations can be adapted to specific tasks by training an additional linear layer,
using predefined hyperparameters tailored for each task.
This task evaluates the adaptability and effectiveness of the pre-trained models when applied to new,
task-specific contexts without altering the original model parameters.

\textbf{Track B: Unparameterized Evaluation.} Pre-trained model embeddings will be used directly for K-nearest neighbor (KNN) classification without training.
This track aims to evaluate the inherent quality of the audio representations without any fine-tuning.
While this approach may not always yield the highest performance in real-world applications,
it serves as a rigorous test of the fundamental representational power of the embeddings.
By avoiding parameterized layers, this track provides a clear view of how well the model captures essential features of the audio data.

\subsection{Training Dataset}

The challenge places a significant emphasis on data collection and utilization, which is a crucial component of the competition.
The organizers do not prescribe a specific training dataset.
Instead, participants are free to use any data for training, as long as it meets the following conditions:

\begin{itemize}[topsep=1ex]
\setlength{\parskip}{0pt}
    \item All training data must be publicly accessible.
    \item Data in Table.\ref{tab:testset} must be excluded from training.
\end{itemize}

\subsection{Datasets for Fine-tuning and Evaluation}

The datasets outlined in Table \ref{tab:testset} encompass a diverse range of audio data across multiple domains, including human voice, environmental sounds, and music. We utilize each dataset's native train-test split to fine-tune and evaluate the participant-submitted models. All datasets are open-source, except for six hidden datasets, detailed in Table \ref{tab:organizer_datasets}, which focus on real-world industrial scenarios provided by the challenge organizers and are marked with $^{\dagger}$ in Table \ref{tab:testset}.

\section{Registration}

To participate, registration is required. Complete the registration form, accessible at \footnote{\url{https://forms.gle/VGgRQdPLs9f72UM8A}}, by the registration deadline of \textbf{April 1, 2025}. Note that this does not means the challenge starts on April 1. The challenge begins on February 7, 2025.

\section{Submission Guide}

Participants are required to submit a pre-trained model encapsulated within the specified API.
The model should accept a single-channel audio signal, represented as a PyTorch tensor with shape $[B, T]$, where $B$ denotes the batch size and $T$ represents the number of samples in the time domain.
The model should output a frame-level prediction of shape $[B, T', D]$, where $T'$ can be different from the input $T$ and $D$ is the embedding dimension defined by the participant.

While there are no strict limitations on model size, submitted models must be able to be run successfully in a Google Colab T4 environment, where the runtime is equipped with a 16 GB NVIDIA Tesla T4 GPU, 12GB RAM.

Participants are also required to submit a technical report along with their submission.

The submission steps are as follows:
\begin{enumerate}
\setlength{\parskip}{0pt}
    \item Clone the audio encoder template from the GitHub repository\footnote{\url{https://github.com/jimbozhang/xares-template.git}}.
    \item Implement your own audio encoder following the instructions in \code{README.md} within the cloned repository. The implementation must pass all checks in \code{audio\_encoder\_checker.py}provided in the repository.
    \item Before the submission deadline, email the organizers \footnote{\url{2025icme-aecc@dataoceanai.com}} a ZIP file containing the complete repository. Additionally, please attach a technical report paper (PDF format) not exceeding 6 pages describing your implementation. Pre-trained model weights can either be included in the ZIP file or downloaded automatically from external sources (e.g., Hugging Face) during runtime. If choosing the latter approach, please implement the automatic downloading mechanism in your encoder implementation.
\end{enumerate}

\section{Evaluation and Ranking}

The performance metrics for each task are normalized to a 0-1 scale, and the final score is computed based on these normalized metrics.

\subsection{Normalization of Metrics}

Each task in Table \ref{tab:testset}, i.e. $ T_i $, has an associated metric $ M_i $ (e.g., accuracy, EER, mAP, F1). To normalize these metrics, we use the following formula:

\begin{equation}
    \hat{M}_i = \frac{M_i - M_i^{\text{min}}}{M_i^{\text{max}} - M_i^{\text{min}}}
\end{equation}
where $ \hat{M}_i $ is the normalized metric for task $ T_i $, and $ M_i $ is the raw metric value for task $ T_i $.
$ M_i^{\text{min}} $ and $ M_i^{\text{max}} $ are the worst and best possible values of the metric $ M_i $, respectively.

For instance, the accuracy, EER, and F1 scores range from 0 to 1, so their $ M_i^{\text{min}} $ and $ M_i^{\text{max}} $ are 0 and 1, respectively; mAP ranges from 0 to 100, so for mAP tasks, $ M_i^{\text{min}} = 0 $ and $ M_i^{\text{max}} = 100 $.

\subsection{Final Score and Ranking}

The final score for each participant for Track A and Track B is calculated as the weighted average of the normalized metrics across all tasks applicable to the respective task,
where the weight is determined by the size of the test set for each task.
This approach ensures that tasks with larger test sets have a greater impact on the final score, reflecting their significance in evaluating the model's performance.
The final scores $S_A$ and $S_B$ for Track A and Track B are given by:

\begin{equation}
    S_{\text{track}} = \frac{\sum_{i=1}^{N_{\text{task}}} n_i \hat{M}_i}{\sum_{i=1}^{N_{\text{task}}} n_i}
\end{equation}
where $N_{\text{task}}$ is the total number of tasks applicable to the respective task,
$n_i$ is the size of the test set for task $T_i$,
and $\hat{M}_i$ is the normalized metric for task $T_i$.

Participants are ranked within each track based on their final scores, $S_A$ and $S_B$, respectively.
The overall performance of the participants will be showcased in two separate leaderboards,
one for Track A and one for Track B, to accurately reflect competencies in both parameterized and unparameterized evaluation methodologies.

\section{Challenge Organizers}

This Challenge is organized by teams from three institutions: Xiaomi Corporation, the University of Surrey and Dataocean AI Inc.

\textbf{Xiaomi Corporation} is a renowned technology company established in 2010. It is widely known for its diverse product range including smartphones, cars, tablets, laptops, wearables, and smart home devices, to form a platform of more than 800 million active devices. The company emphasizes innovation and user experience, is dedicated to fundamental technologies, blends into open-source.
AI has been fully integrated into to reinforce Xiaomi's machie intelligence and service efficiency, ranging from user interaction, imaging, auto pilot, to internet sales, delivery, and service. Among them, the acoustic and speech team of the AI lab is committed to us large audio and speech models to boost the research and development in speech recognition, speech synthesis, microphone array based noise reduction, voice trigger, extraction and understanding of rich language, and acoustic measurement.

\textbf{Dr. Junbo Zhang} is an AI Research Scientist at Xiaomi Corporation. He earned his Ph.D. from the Institute of Acoustics at the Chinese Academy of Sciences. With years of experience in developing acoustic and speech algorithms, Dr. Zhang has made significant contributions to various fields, including speech recognition, pronunciation evaluation, speech synthesis, audio tagging, sound separation, and noise reduction.
He has authored over 30 papers in prestigious journals and top-tier conferences. As a code contributor to the open-source project Kaldi, he also wrote the book ``Kaldi Speech Recognition Practice'',which has sold more than ten thousand copies.
At Xiaomi, he was instrumental in developing and launching the company's initial speech recognition system, the wake word detection for ``Xiao Ai'' (Xiaomi's AI assistant), and the voiceprint recognition system. Currently, he leads several pioneering projects in the large model technology domain, pushing the boundaries of what is possible in consumer electronics.

\textbf{University of Surrey} The Machine Audition Lab within the Centre for Vision Speech and Signal Processing at the University of Surrey, led by Prof Wenwu Wang, is a leading research lab in audio signal processing and machine learning, consisting more than 30 researchers. They have developed several widely used audio representation models such as PANNs, AudioLDM, AudioLDM 2, AudioSep, etc. They have been contributing to the activities in Detection and Classification of Acoustic Scenes and Events (DCASE) challenges and workshops since 2013, including the organisation of two tasks of the DCASE 2024 Challenges, i.e. Task 6 - Automated Audio Captioning and Task 9 - Language-Queried Audio Source Separation. 

\textbf{Dr. Wenwu Wang} is a Professor in Signal Processing and Machine Learning, University of Surrey, UK. He is also an AI Fellow at the Surrey Institute for People Centred Artificial Intelligence. His current research interests include signal processing, machine learning and perception, artificial intelligence, machine audition (listening), and statistical anomaly detection. He has (co)-authored over 300 papers in these areas. He has been recognized as a (co-)author or (co)-recipient of more than 15 accolades, including the 2022 IEEE Signal Processing Society Young Author Best Paper Award, ICAUS 2021 Best Paper Award, DCASE 2020 and 2023 Judge’s Award, DCASE 2019 and 2020 Reproducible System Award, and LVA/ICA 2018 Best Student Paper Award. He is an Associate Editor (2020-2025) for IEEE/ACM Transactions on Audio Speech and Language Processing, and an Associate Editor (2024-2026) for IEEE Transactions on Multimedia. He was a Senior Area Editor (2019-2023) and Associate Editor (2014-2018) for IEEE Transactions on Signal Processing. He is the elected Chair (2023-2024) of IEEE Signal Processing Society (SPS) Machine Learning for Signal Processing Technical Committee, a Board Member (2023-2024) of IEEE SPS Technical Directions Board, the elected Chair (2025-2027) and Vice Chair (2022-2024) of the EURASIP Technical Area Committee on Acoustic Speech and Music Signal Processing, an elected Member (2021-2026) of the IEEE SPS Signal Processing Theory and Methods Technical Committee. He has been on the organising committee of INTERSPEECH 2022, IEEE ICASSP 2019 \& 2024, IEEE MLSP 2013 \& 2024, and SSP 2009. He is Technical Program Co-Chair of IEEE MLSP 2025. He has been an invited Keynote or Plenary Speaker on more than 20 international conferences and workshops.

\textbf{Dataocean AI Inc.} is a global data collection and labeling services provider that combines technology with a diverse network of millions data contributors, scientists, and engineers. The company delivers cutting-edge data solutions across multiple domains, including text, audio, image, and multimodal for foundation models or GenAI applications. With over 1,600 off-the-shelf datasets and a proven track record of delivering thousands of customized data projects, Dataocean AI have been trusted by of over 1,000 global AI leading enterprises and institutions. The company cover more than 200 languages around the world. Its self-developed data platform ensures precision and efficiency in tasks such as collection, cleansing, labeling and evaluation.  With nearly two decades of experience, Dataocean AI has established itself as a trusted partner in the AI ecosystem, consistently delivering excellence and earning global recognition.

\section{Challenge Schedule}

The Challenge will follow this schedule:

\begin{itemize}[topsep=1ex]
\setlength{\parskip}{0pt}
    \item February 7, 2025: Challenge announcement.
    \item April 30, 2025: Submission deadline.
    \item May 27, 2025: Results announcement.
\end{itemize}

\bibliographystyle{IEEEtran}
\bibliography{paper}   

\appendix
\appendixpage
\addappheadtotoc

\section{Related Work}

Our challenge broadens the scope by including non-speech related tasks, enabling a more comprehensive evaluation of audio encoders, which are pivotal in both continuous and discrete audio processing contexts.
Here, we discuss three of the existing benchmarks, highlighting the unique contributions and improvements of our proposed challenge.

\subsection{HEAR: Holistic Evaluation of Audio Representations}
Our proposed challenge is strongly inspired by the HEAR benchmark~\cite{turian2022hear}, which assesses audio representations across environmental sound and music tasks.
While HEAR provides an excellent foundation, our challenge introduces several enhancements:

\textbf{Diverse task set:}  HEAR comprises 19 tasks in total, 17 of which are unique, while two tasks differ only in their available training data.
While the tasks in HEAR encompass various application scenarios for sound event detection and music processing, they lack variety in human voice processing.
Our challenge offers a more comprehensive and balanced distribution of tasks across human voice, music, and environmental sound domains,
leveraging a suite of open-source datasets that reflect real-world scenarios and user experiences,
including unique datasets such as car scratching, inside/outside car environments.

\textbf{Focus on real-world applications:}
Some tasks in HEAR, although interesting, may have limited applications and high variance during testing (e.g., Gunshot Triangulation and Beehive) due to the factors such as small sample sizes, which have led to many follow-up works discarding those tasks.
We seek to balance task variety, real-world impact, and robust performance estimation, ensuring the evaluated representations are relevant to industrial use, providing reliable performance metrics.

\textbf{Evaluation methods:}
In addition to linear projection, we utilize unparameterized methods for classification.
This evaluation aims at investigating the use of features for cases such as unsupervised clustering.

\textbf{Efficient system:} We propose an open-sourced, efficient evaluation system, incorperating a simple pipline that can be run without any prequisites.


\subsection{SUPERB: Speech processing Universal PERformance Benchmark}
SUPERB~\cite{yang2021superb} and its derivatives primarily focus on speech processing tasks using self-supervised learning (SSL) representations.
In recent years, SUPERB also included additional tasks such as emotion recognition and sound codecs, but notably, it does not include environmental audio or music related tasks.

Our challenge broadens this scope with the inclusion of non-speech related tasks (environmental audio, music), enabling a more comprehensive evaluation of audio representations.

\subsection{DASB: Discrete Audio and Speech Benchmark}
DASB~\cite{mousavi2024dasb} benchmarks discrete audio tokens across various tasks, mainly focuses on the speech domain.
While discretization is an important research field, continuous representations offer complementary advantages.
Continuous representations directly addresses the need for robust audio encoders in multimodal applications, where continuous embeddings are often preferred for seamless integration and efficient processing \cite{wang2024comparative,yu2024salmonn}.
The output of this challenge can be used to complement discrete representation research by, for example, injecting general semantic information into codecs \cite{kyutai2024moshi},
or evaluating the loss of information during the discretisation process.
Our challenge prioritizes established methods for continuous representations.

\end{document}